# Hysteretic magnetoresistance in nanowire devices due to stray fields induced by micromagnets

Y. Jiang[1], E.J. de Jong[2], V. van de Sande[2], S. Gazibegovic[2], G. Badawy[2], E.P.A.M. Bakkers[2], S.M. Frolov[1] *

[1]*University of Pittsburgh, Pittsburgh, Pennsylvania 15260, USA*
[2]*Eindhoven University of Technology, 5600 MB, Eindhoven, Netherlands*
* frolovsm@pitt.edu

**Abstract**

We study hysteretic magnetoresistance in InSb nanowires due to stray magnetic fields from CoFe micromagnets. Devices without any ferromagnetic components show that the magnetoresistance of InSb nanowires commonly exhibits either a local maximum or local minimum at zero magnetic field. Switching of microstrip magnetizations then results in positive or negative hysteretic dependence as conductance maxima or minima shift with respect to the global external field. Stray fields are found to be in the range of tens of millitesla, comparable to the scale over which the nanowire magnetoresistance develops. We observe that the stray field signal is similar to that obtained in devices with ferromagnetic contacts (spin valves). We perform micromagnetic simulations which are in reasonable agreement with the experiment. The use of locally varying magnetic fields may bring new ideas for Majorana circuits in which nanowire networks require control over field orientation at the nanoscale.

**Introduction**

InSb nanowires demonstrate ballistic transport [1] and strong spin-orbit interaction [2]. These properties make them promising for the realization of Majorana bound states (MBS) in hybrid semiconductor-superconductor structures. In Majorana experiments that are typically performed in an external global magnetic field, transport measurements show zero bias conductance peaks, which are investigated as signatures of MBS [3]. Magnetic materials may replace the global magnetic field in MBS experiments and provide additional design and architecture freedom for Majorana-based quantum circuits [4-6]. It may be possible to turn off the external magnetic field after the micromagnets are magnetized in a desired configuration. The stray field profile may in principle reproduce a helical pattern of spin-orbit field and therefore broaden the material selection for MBS experiments to those without intrinsic spin-orbit interaction [7]. The usefulness of micromagnets for quantum devices is not limited to Majorana experiments, as micromagnets have been used in quantum dot and spin qubit experiments [8-10].

In this article, by integrating ferromagnetic components into nanowire devices, we show that the local stray magnetic field can influence conductance of InSb nanowires. For our most conclusive experiments, we fabricate InSb nanowire devices with non-magnetic ohmic contacts. Two ferromagnetic microstrips with different dimensions are deposited beside the InSb nanowire but without forming ohmic contacts. The coercivity of microstrips and stray field inside the nanowire can be estimated from switches of magnetoresistance and shift of the local magnetoresistance maximum or minimum. The coercivity difference of the two ferromagnetic microstrips is of order 10mT, and the stray field within the nanowire is of order tens of millitesla. We also perform magnetoresistance measurements on devices in which the ferromagnets do make ohmic contact to the nanowire. We observe similar hysteresis which can be fully explained by the local stray field effects. In future work, it may be feasible to dispense with the external magnetic field in MBS experiments. However, in devices studied here the main direction of the stray field produced by the ferromagnetic micro-strips is not along the nanowire, which is needed for the realization of MBS. By careful design and arrangement of ferromagnetic microstrips, a required profile of local stray field can be generated [5].

**Device Fabrication Methods**

We distinguish between ferromagnetic strips in ohmic contact with the nanowire which we denote as F, and similar strips not in ohmic contact with the nanowire which we denote as (F). We fabricate four types of devices for our experiments: N-(F,F)-N; N-N; N-F-F-N; and F-F. For N-(F,F)-N devices, there are two non-magnetic ohmic contacts labeled N and two ferromagnetic microstrips (F, F) not in ohmic contact with the nanowire. For N-F-F-N and F-F devices, there are ferromagnetic ohmic contacts. There are two types of nanowires used and for more details of the nanowires, we refer to growth papers [11, 12]. Long stemless InSb nanowires are used for N-(F,F)-N, N-N and N-F-F-N devices; short stemmed nanowires are used for F-F devices. Nanowires are placed on a back-gate chip, enabling chemical potential tuning.

There are two types of back gate chips used in our experiment. The first type is using highly doped silicon with a layer of thermal silicon oxide and a layer of 10nm dielectric HfOx deposited by atomic layer deposition (ALD). For the second type, a thin Ti/Au (1nm/10nm) layer is deposited on an undoped silicon chip. A dielectric HfOx layer (10 nm) is deposited using ALD on top of the metal layer. This second type of gate yields improved charge stability of nanowire devices. Both gates tune the entire nanowire. To fabricate ferromagnetic ohmic contacts, we use sulfur passivation to remove native oxide before the deposition of CoFeB (F) by electron beam evaporation. The electron beam evaporator cannot preserve the atomic ratio of CoFeB pellets (30/55/15 before deposition) and the deposited metal is expected to be CoFe [13]. Before the deposition of Ti/Au (non-magnetic contacts, N), argon ion milling is used to remove the native oxide.

For N-(F,F)-N devices, we emphasize that there are only Ti/Au ohmic contacts formed and the ferromagnetic microstrips are intentionally deposited either beside the nanowires or on top of nanowires but without removing the native oxide layer thus preventing ohmic contact. In contrast, for N-F-F-N and F-F devices, sulfur passivation is performed for ferromagnetic contacts. Another fabrication consideration to point out is the choice of thickness for CoFe. For N-F-F-N and F-F devices, to overcome the thickness of the nanowire (typically around 100nm), the thickness of CoFe is around 70nm and we perform deposition at a finite angle (typically $50°$ with respect to the chip surface). For N-(F,F)-N devices, since we do not need to cover the nanowire, the thickness of the CoFe strips is thinner: around 40 nm. Our micromagnetic simulations indicate that this smaller thickness may facilitate sharper magnetization switching because the strip is closer to the single domain regime.

**Experimental Results**

We demonstrate the spin valve-like signal due to the stray magnetic fields first using N-(F,F)-N devices (Fig. 1(a)). The measurement is performed in a dilution refrigerator with a base temperature of 40 mK. For this measurement, we apply a DC voltage bias across the non-magnetic N contacts and measure current through the nanowire, as shown in Fig. 1(b). Magnetic field is applied parallel to the easy axis of the CoFe strips, in the case of device A presented in Fig. 1 there is a 70° angle between the direction of the field and the nanowire. The two CoFe microstrips are expected to have different coercivities because of their different widths, so the magnetoresistance of the device is expected to exhibit two sharp switches for each magnetic field orientation. Fig. 1(c) shows a representative measurement. Indeed, there are hysteretic switches at -42mT, -63mT, 43mT, and 53mT. We also notice that the current maximum (minimum of magnetoresistance) is shifted away from zero field. For the red curve it is at -25mT and for the blue curve it is at 35mT. Thus, the local stray field in this measurement is on the order of 30 mT. For different scans, these values vary but remain similar (see supplementary materials). The error of determining switch positions is ± 1mT from scan resolution.

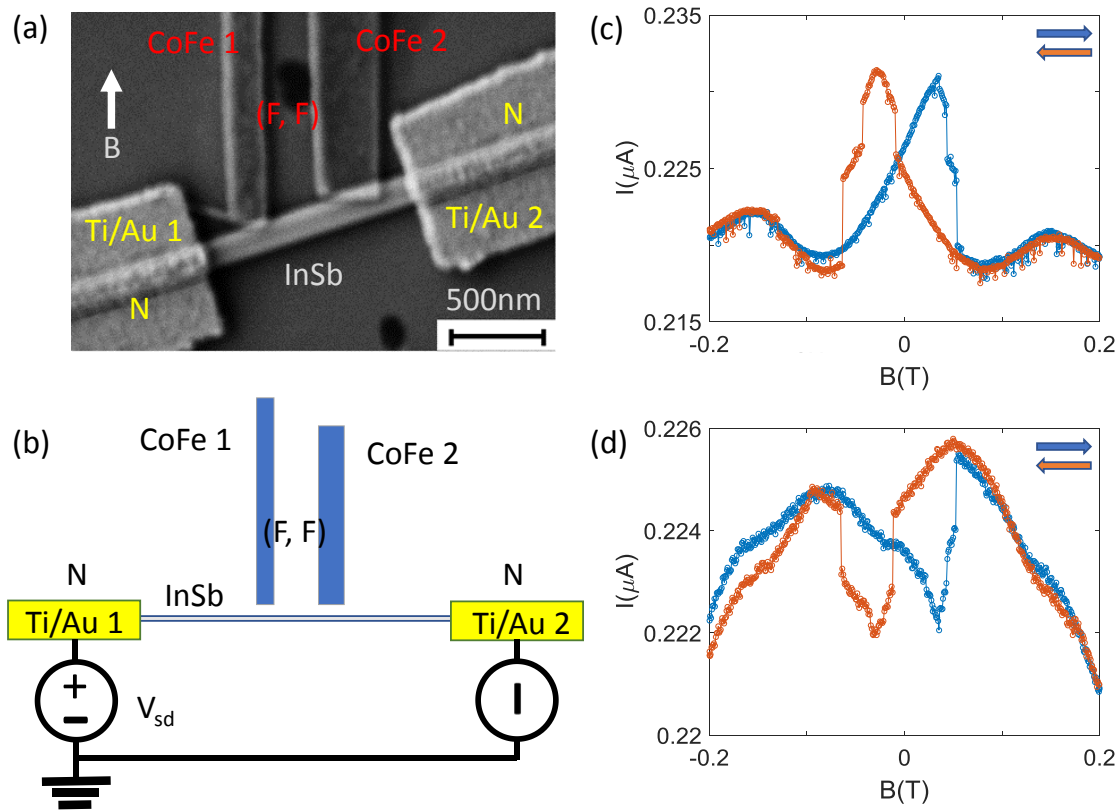

**Figure 1**. (a) Scanning electron microscope (SEM) image of a N-(F,F)-N device A. The two CoFe strips are insulated from the nanowire by native oxide. The direction of the external magnetic field B is indicated by arrow. (b) Schematic of measurement setup. (c) Current as function of external magnetic field using the source-drain bias voltage $V_{sd} = 2mV$ and gate voltage $V_g = 1.05V$. Directions of magnetic field sweeps are indicated by arrows. (d) Magnetic field scan taken at $V_{sd} = 2mV$ and $V_g = 1.1V$.

By adjusting the gate voltage or bias voltage, it is possible to change the sign of the hysteresis, as shown in Fig. 1(d). The magnetoresistance of InSb nanowires may have a local maximum or a local minimum at zero magnetic field. Figs. 1(c,d) show a transition from minimum to maximum resistance resulting in sign change of the hysteresis. We need to point out that conductance jumps due to local charge rearrangement are also commonly observed. To distinguish magnetic jumps from charge jumps, we look for approximate symmetry around zero field and for approximate reproducibility from sweep to sweep of the external magnetic field. In Fig. 1(d), the red and blue curves do not overlap in the range from -0.2T to -0.1T, while they do overlap in the positive field direction. This may be due to an irreproducible charge jump during negative field sweep. We present more hysteretic magnetoresistance plots to justify that the observed hysteresis is not from charge jumps in supplementary materials.

In supplementary materials we also present magnetoresistance measurements on a device without any ferromagnetic components (N-N). The magnetoresistance of that device shows a dip feature around zero magnetic field and no hysteresis. This confirms that the hysteretic magnetoresistance originates from switching in micromagnets and that the electromagnet used to generate the global external magnetic field is not hysteretic.

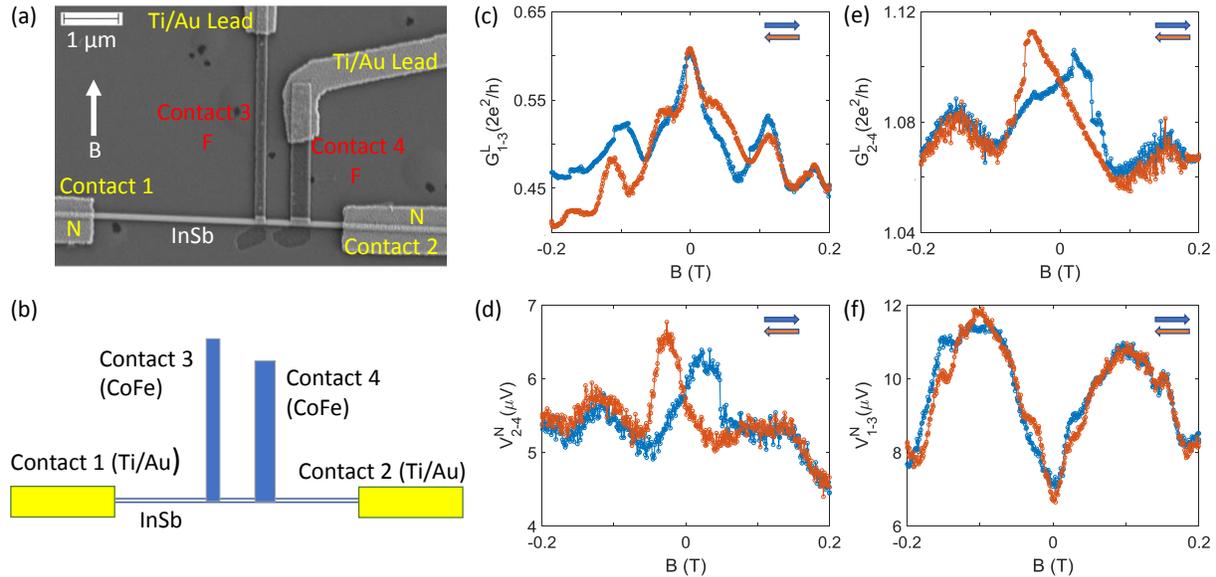

**Figure 2**. (a) Scanning electron microscope (SEM) image of N-F-F-N device B. The direction of the applied magnetic field B is indicated by arrow. (b) Schematic of device B. Data shown in this figure are acquired by a lock-in. (c, d) Local and non-local signals from magnetic field scan taken at 0.1mV AC voltage bias, zero DC voltage bias and $V_g = 1.2V$ in configuration 1. (e, f) Local and non-local signals from magnetic field scan in configuration 2 taken at 0.1mV AC voltage bias, 1 mV DC voltage bias and $V_g = 1.35V$. For (c) and (e), the resistance of the measurement circuit is not subtracted (around 4.5kΩ). For (d) and (f), the voltage is the AC voltage measured by lock-in.

We find similar hysteretic signals in N-F-F-N devices such as device B shown in Figs. 2(a,b). The geometry of the CoFe contacts is close to that in device A, but now the ferromagnets are in ohmic contact with the nanowire. The four-terminal contact geometry allows us to measure local current and non-local voltage simultaneously, as in the non-local spin valve experiments. The direction of the external magnetic field is 5 degrees away from perpendicular to the nanowire. We use two measurement configurations. First, we apply voltage bias and measure local conductance across the left side (contacts 1 and 3). The non-local voltage is measured across the right side (contacts 2 and 4). In this configuration presented in Figs. 2(c,d) we observe hysteretic signals non-locally without observing a clear local hysteresis. The maximum of local signal is at zero external field. Such a pattern is expected from a basic non-local spin valve where only spin-polarized transport in between the two ferromagnets should exhibit hysteretic switching. However, measurements in the second configuration are inconsistent with such basic non-local spin valve picture. Voltage bias is now applied across the right side (contacts 2 and 4). Using this setup, we observe hysteresis in the local conductance (Figs. 2(e,f)). In the non-local voltage signal, the minimum of the signal is centered at zero field, though small hysteretic loops appear symmetrically around zero field as well. In the supplementary materials we show more data for both measurement setups. It should be pointed out that plots without clear hysteresis are common which may be due to charge jumps and less sharp magnetization switches in this device. Nevertheless, the hysteretic signals in device B are consistent with the hysteretic signals in device A in terms of their extent in magnetic field. Also consistent with measurements on device A in Fig. 1, the sign of the hysteresis can also be modulated by adjusting gate voltages and bias voltages in device B, in contradiction to the simple spin valve picture in which a lower non-local signal is expected for antiparallel magnetization configuration. The CoFe contacts of device B are expected to produce similar local stray fields as in device A. Therefore, we conclude that the stray field is more likely the origin of the observed hysteresis as opposed to spin-polarized transport.

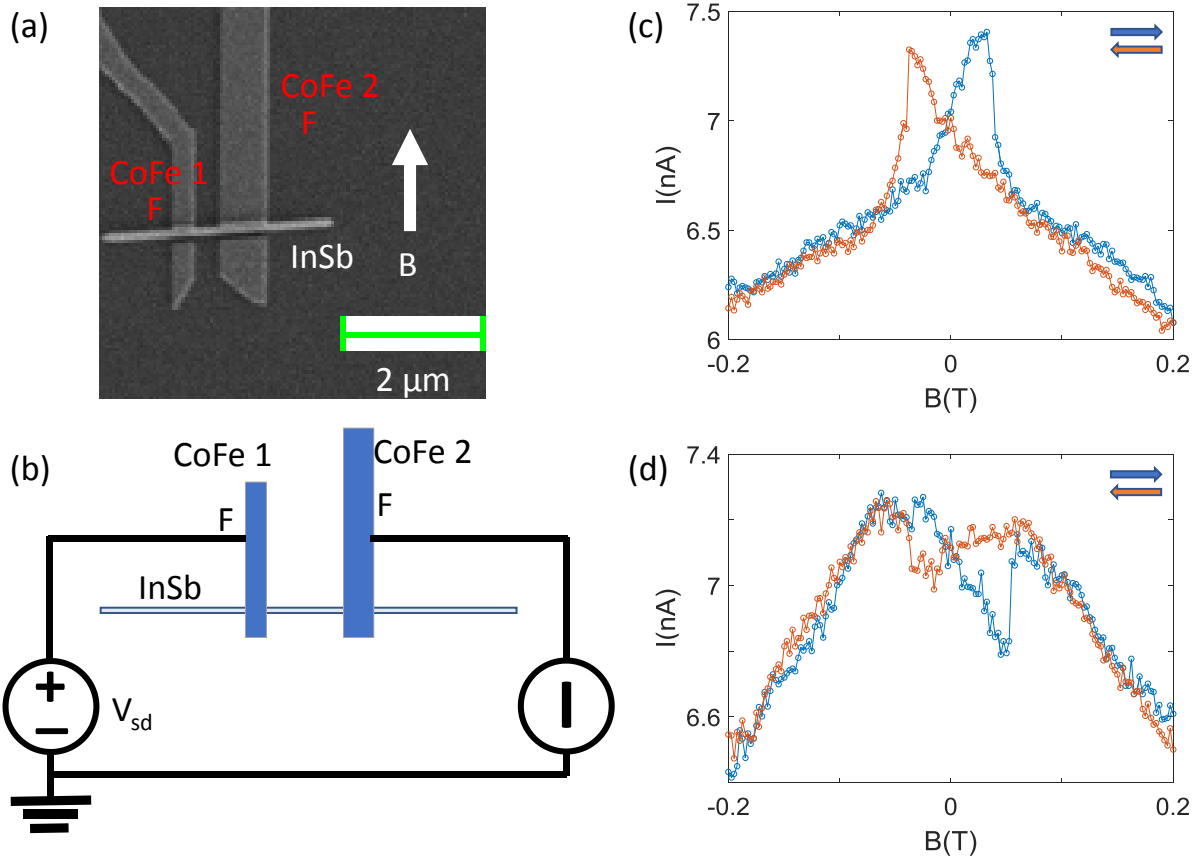

**Figure 3**. (a) Scanning electron microscope (SEM) image of an F-F device C. The direction of the applied magnetic field B is indicated by arrow. (b) Schematic of measurement setup. All data shown in this figure are acquired by DC measurement. (c) Current as a function of magnetic field taken at $V_{sd}$ = 0.1mV and $V_g$ = 0V. (d) Magnetic field scan taken at $V_{sd}$ = 0.1mV and $V_g$ = 0.08V. The sign of the hysteresis is reversed compared to panel (c).

As additional experiments, we perform DC measurements using devices with only two ferromagnetic contacts (F-F devices), which is commonly known as the local spin valve geometry[14]. One of these devices is shown in Fig. 3(a) and the measurement setup is shown in Fig. 3(b). Figs. 3(c,d) show two representative magnetic field scans. The features of the hysteresis are consistent with what we show in Figs. 1 and 2. Thus, also in this local spin valve configuration the hysteretic switching magnetoresistance is consistent with the effect of stray magnetic fields from the microstrips.

**Micromagnetics simulations**

We also perform simulations of stray fields to compare with experimental results. For more realistic results, we use the geometry of device A including the shapes of CoFe strips and the orientation of the nanowire. In the simulations the two CoFe microstrips have dimensions of 150nm × 4500nm × 40nm and 300nm × 3000nm × 40nm respectively. To take into account the small CoFe overlap with the nanowire (Fig. 1(a)), we take the intersection of the two strips and the nanowire and lift this intersection onto the top of the nanowire. Simulations are carried out using MuMax3 [15, 16]. The magnetic parameters of CoFe vary in a broad range [17-20] and the magnetic parameters of our CoFe films are not known. For simulations we therefore use parameters known for Co which we expect to give values close to CoFe. We set the magnetic saturation to be $1.44 \times 10^6$ A/m, the first order uniaxial anisotropy constant to be $4.1 \times 10^{-1}$ J/m$^3$, the second order uniaxial anisotropy constant to be $1.4 \times 10^{-1}$ J/m$^3$, the exchange constant to be $3 \times 10^{-11}$ J/m, and the phenomenological damping constant to be 0.01. We use a 10nm cubic cell size in the interest of reducing computation time, though the exchange length of CoFe is likely smaller than 10nm.

Fig. 4(a) shows the geometry of the micro-strips and the nanowire. The coordinate along the nanowire is u, and v is perpendicular to nanowire. Fig. 4(b) shows the averaged and normalized magnetization in y direction of the strips. The simulations reproduce the experimentally observed double-switch hysteretic loops (Fig. 1(c) and Fig. 4(b)). The anti-parallel configurations appear at around ±50mT, which is consistent with what we observe experimentally. This loop is not symmetric, which indicates that the stray field from the strips may influence the magnetization. We can also observe small switches close to 0 field. This originates from the small regions of ferromagnet on top of the nanowire. In Fig. 4(c), we show the stray field along the nanowire when the strips are magnetized in parallel (+y direction). To generate stray field plots we take the average of the stray field inside the hexagonal nanowire cross section with the size of 100 nm (the thickness of the nanowire). As expected, the field in the v direction (the direction perpendicular to the nanowire) is the largest. However, the components of the field in the u and z directions are comparable. The stray field profile is not uniform due to the lack of symmetry of the two microstrips. The maximum stray field inside the nanowire is close to 150mT. Our experimental data indicate a factor of 2-3 smaller stray field inside the nanowire. This is likely because the saturation magnetization used in our simulation overestimates the experiment.

**Alternative Explanations**

When micromagnets are in ohmic contact with the nanowire, spin injection and pure spin currents can produce hysteretic signals such as in our N-F-F-N and F-F devices [21-23]. While spin injection may be taking place, it is challenging to separate its effect from the effect of local stray fields since both effects manifest in similar fashion (see Figs. 2 and 3). However, it is difficult to explain the hysteretic signal sign reversal with spin injection. Besides directly injecting spin polarized current using ferromagnetic ohmic contacts, there are other methods to induce hysteretic magnetoresistance that use the exchange field from ferromagnetic insulators (or metals) [6] which is expected to have a significantly smaller effect on magnetoresistance than the stray field effect. Another possibility is magneto-Coulomb effect [24, 25] which similarly to the local stray field effect requires only a single ferromagnet to manifest itself.

**Conclusion and Future Work**

Our experiments demonstrate the influence of the stray fields produced by CoFe micromagnets on transport in InSb nanowires. We observe that the stray field effect can mimic spin polarized transport signatures such as hysteretic transport. In future work, by careful design of ferromagnetic microstrips, we shall aim to realize stray field profiles with field along the nanowire and search for MBS in a hybrid superconductor-semiconductor system.

**Further Reading**

General introduction to spintronics and spin valve measurements can be found here [26-29]. A reader may find helpful recent reviews of Majorana physics in nanowires [4, 30-34]. For the readers who are interested in MBS devices without external magnetic field we suggest these articles [5, 35, 36]. The following papers discusses other experiments where micromagnets are combined with nanowires [8, 37]. Experiments on spin qubits in quantum dots with the use of micromagnets are described here [8-10].

**Duration and Volume of Study**

This article is written based on more than 6500 datasets from 8 cool downs of dilution fridges. For each cool down, we measure several devices. We measured 19 N-(F,F)-N devices, 1 N-N device, 20 N-F-F-N devices and 9 F-F devices. Among them, we observed reproducible hysteretic magnetoresistance in 17 devices. The yield of the measurable devices is limited by fabrication and occasional static discharge damage to devices.

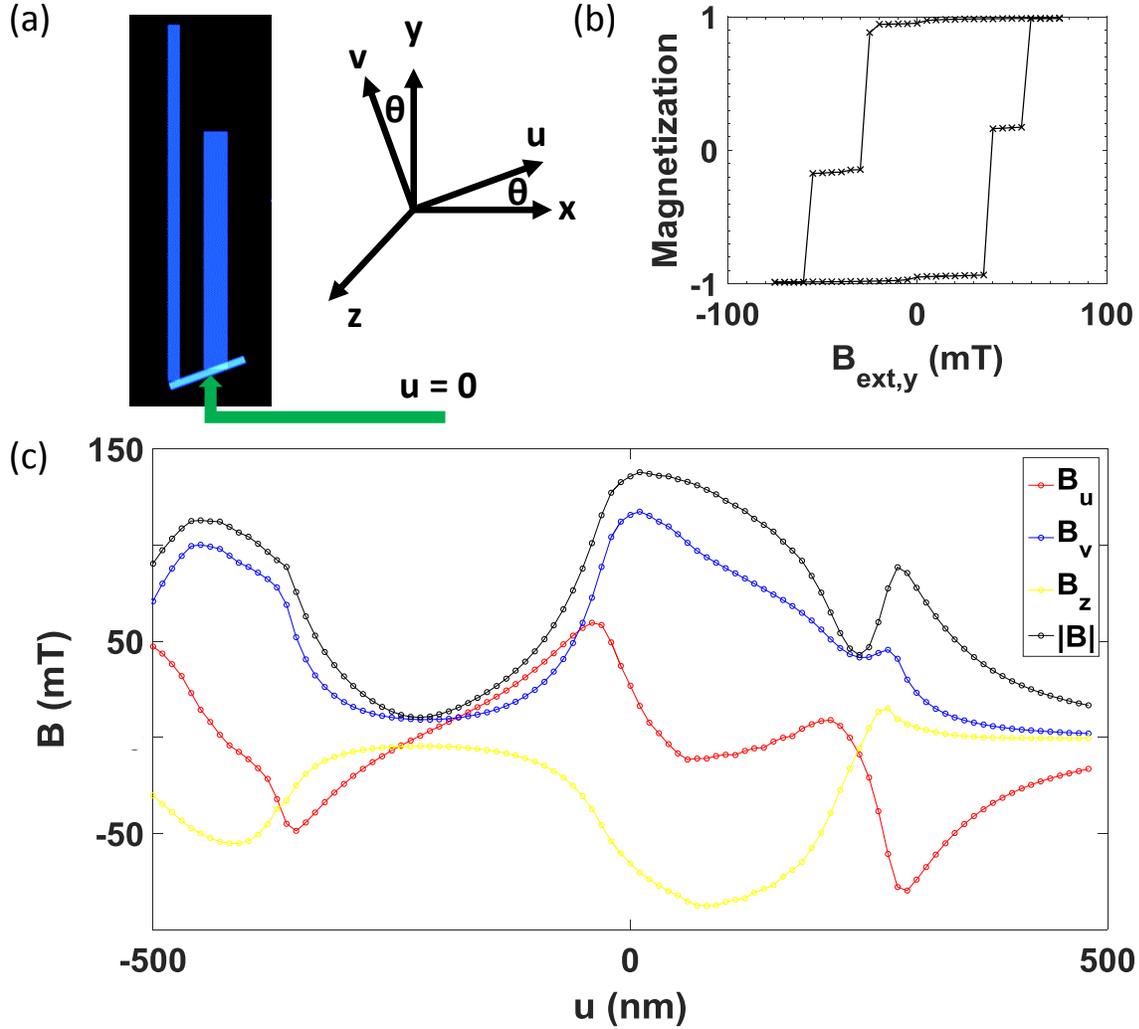

**Figure 4**. (a) The numerical simulation geometry and the coordinate system. The origin of the coordinates is located at the center of the nanowire segment. u axis is along the nanowire. The angle θ between x and u is 20°. (b) Average normalized magnetization in y direction of the two strips. (c) The stray field in the nanowire for parallel up magnetization (achieved when $B_{ext,y}$=75 mT) as function of u.

## Data availability

A curated set of experimental data (including Mumax3 script for micromagnetics simulations) is available at: https://github.com/frolovgroup/Yifan-Jiang. The data and an interactive version of the paper are in the folder named as "Hysteretic magnetoresistance in nanowire devices due to stray fields induced by micromagnets".

## Author Contributions

Y.J. performed device fabrication and measurements. G.B., S.G. and E.B. provided InSb nanowires. Y.J. and S.F. analyzed the experimental data. E.d.J., V.v.d.S. and Y.J. performed micromagnetics simulations. Y.J. and S.F. wrote the manuscript with contributions from all co-authors.

## Acknowledgements

We thank V. Pribiag, Z. Yang, M. Pioro-Ladrière and M. Jardine for discussions. Work supported by the Department of Energy DE-SC-0019274.

# Supplementary Materials

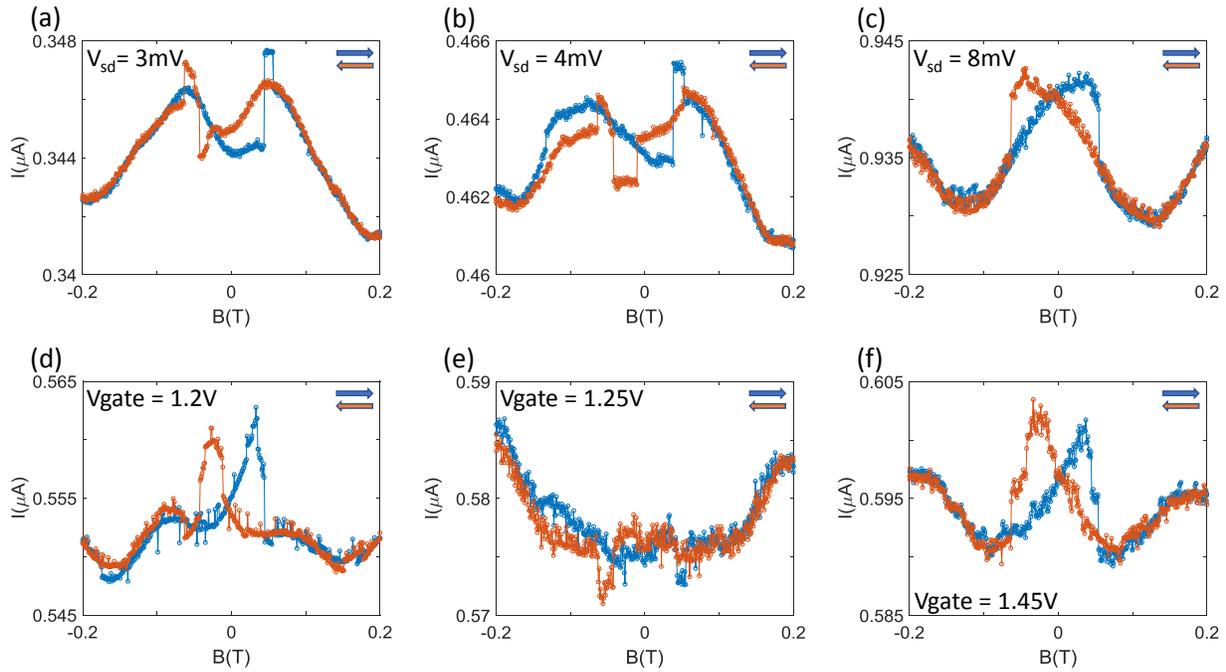

**Figure S1**. Additional magnetic field scans measured in device A, a N-(F,F)-N device. We observe that the response can be modulated by gate voltage and bias voltage. Current vs. magnetic field measured at different gate and bias settings. Data are acquired by DC measurement. (a-c) at $V_g$ = 1.1V. (d-e) at $V_{sd}$ = 5mV.

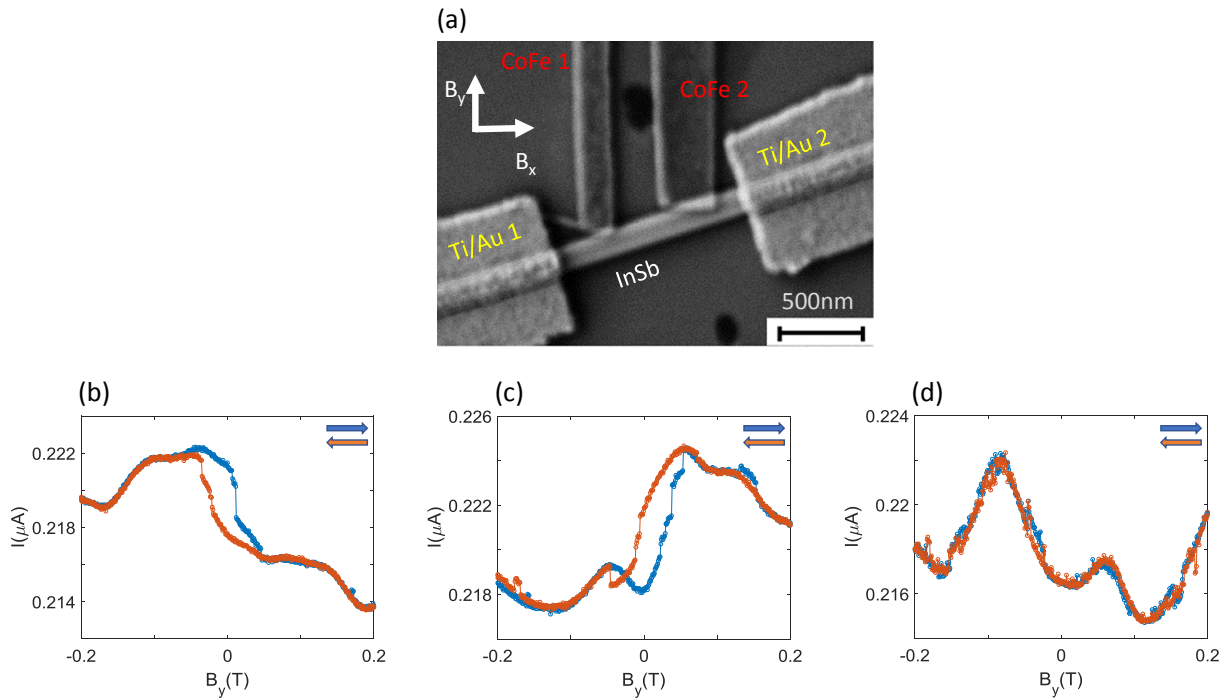

**Figure S2. Measurements in a finite $B_x$** (a) SEM image of device A with field directions indicated (b-c) $B_y$ magnetic field scan with $B_x = -0.2T$ and $B_x = 0.2T$ respectively. Data are acquired by DC measurement. Both plots are measured at $V_{sd} = 2mV$ and $V_g = 1.05V$. Magnetoresistance is not symmetric, but hysteresis is observable. (d) $B_y$ magnetic field scan with $B_x = -1T$. The measurement is performed at $V_{sd} = 2mV$ and $V_g = 1V$. At this large $B_x$ field, there is no observable hysteresis as the magnetization of CoFe strips is pinned by $B_x$. Measurements are performed using a 2D vector magnet with sample plane aligned with magnetic field plane.

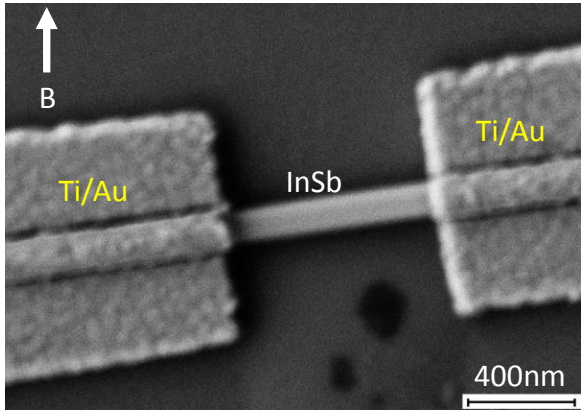
(a)

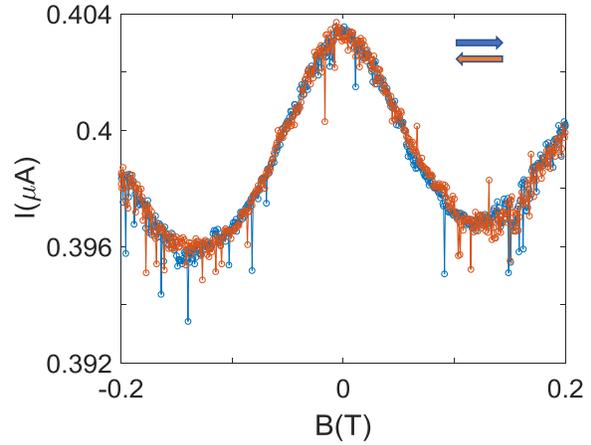
(b)

**Figure S3.** Measurement on N-N device (a) Scanning electron microscope image of device D, a device without any ferromagnetic components. (b) Representative magnetic field scan taken at $V_{sd} = 5mV$ and $V_g = 1V$. Data are acquired by DC measurement. There is no observable hysteresis and the signal is peaked at zero field. The peak width is of order 100 mT.

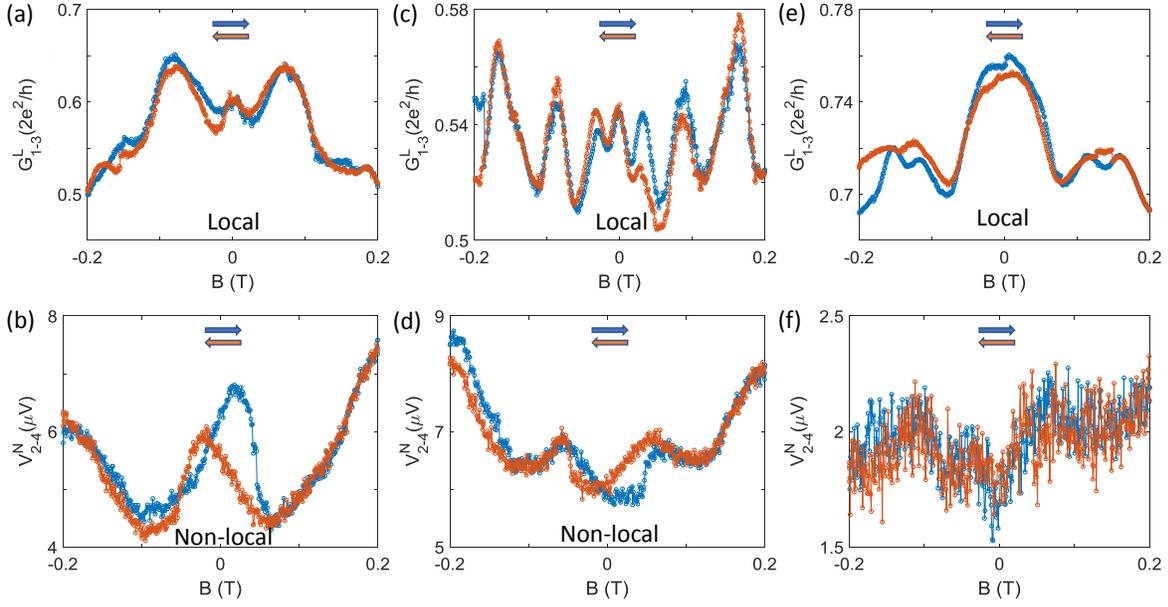

**Figure S4**. Additional data from N-F-F-N device. Magnetic field scan measured from device B shown in Fig. 2 (a). Voltage is applied across contacts 1 and 3. Non-local signal is the voltage across contacts 2 and 4. (a,b) Local and non-local magneto-conductance measured at 0.1mV AC bias and 1.2V gate voltage. (c,d) Local and non-local magneto-conductance measured at the same bias and gate setting as in (a,b), but at a different time. (e,f) Local and non-local magneto-conductance measured at 0.1mV AC bias and 2.85V gate voltage.

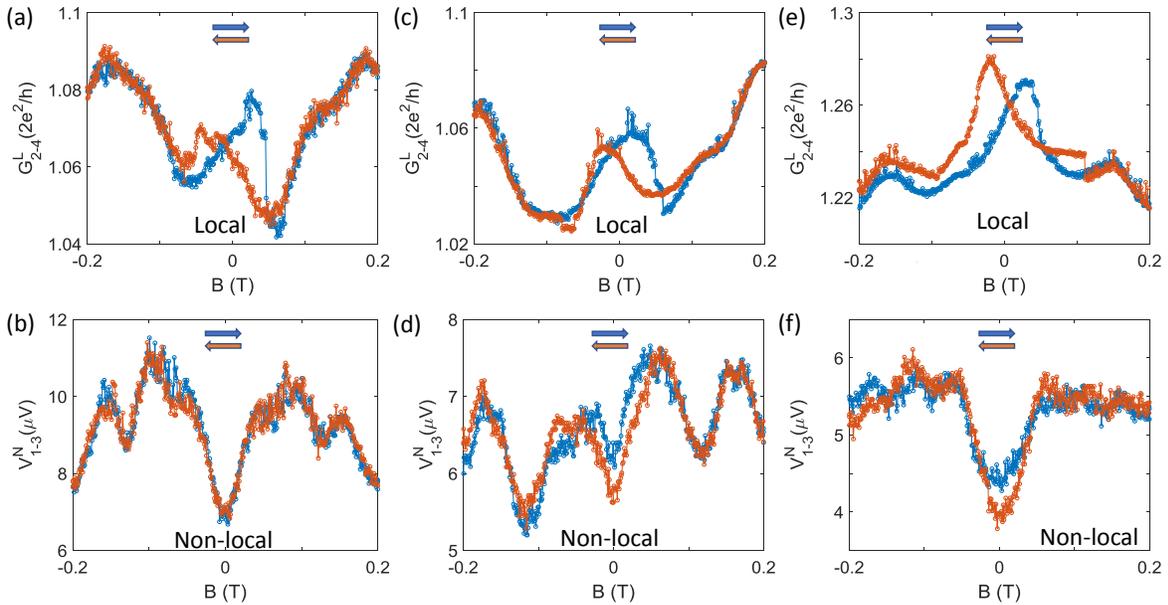

**Figure S5**. More data from N-F-F-N Device B. (a)Voltage is applied across contacts 2 and 4. Non-local signal is the voltage across contacts 1 and 3. (a,b) Local and non-local magneto-conductance measured at 1mV DC bias, 0.1mV AC bias and 1.35V gate voltage. (c,d) Local and non-local magneto-conductance measured at 1mV DC bias, 0.1mV AC bias and 1.5V gate voltage. (e,f) Local and non-local magneto-conductance measured at 1mV DC bias, 0.1mV AV bias and 2.7V gate voltage.

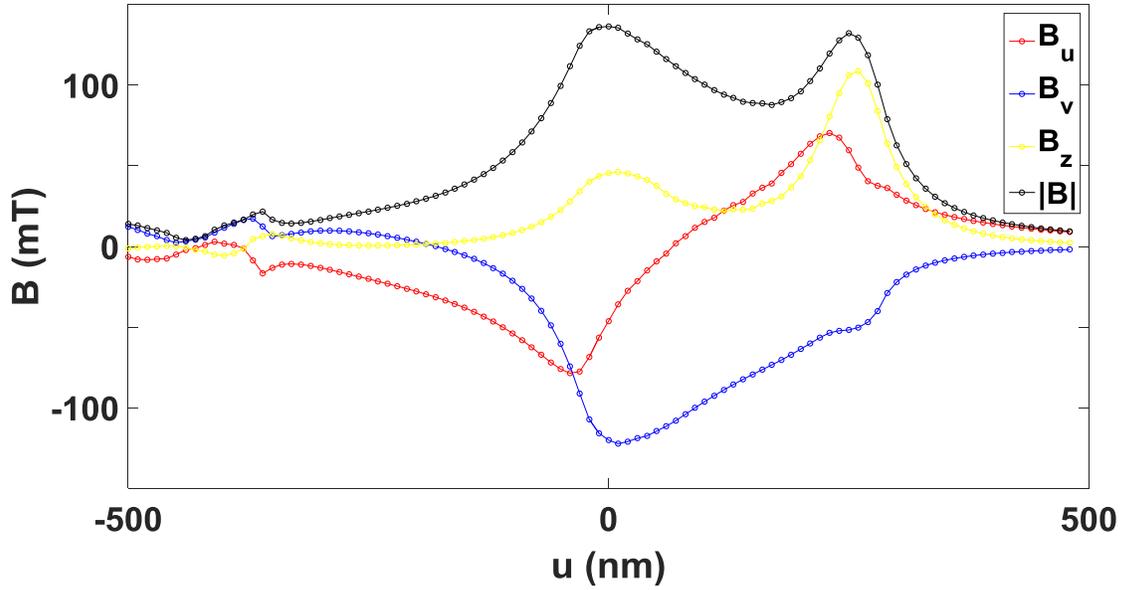

**Figure S6**. The averaged stray field in the nanowire for anti-parallel up-down magnetization ($B_{ext,y} = -50\ mT$) as function of the nanowire coordinate u.

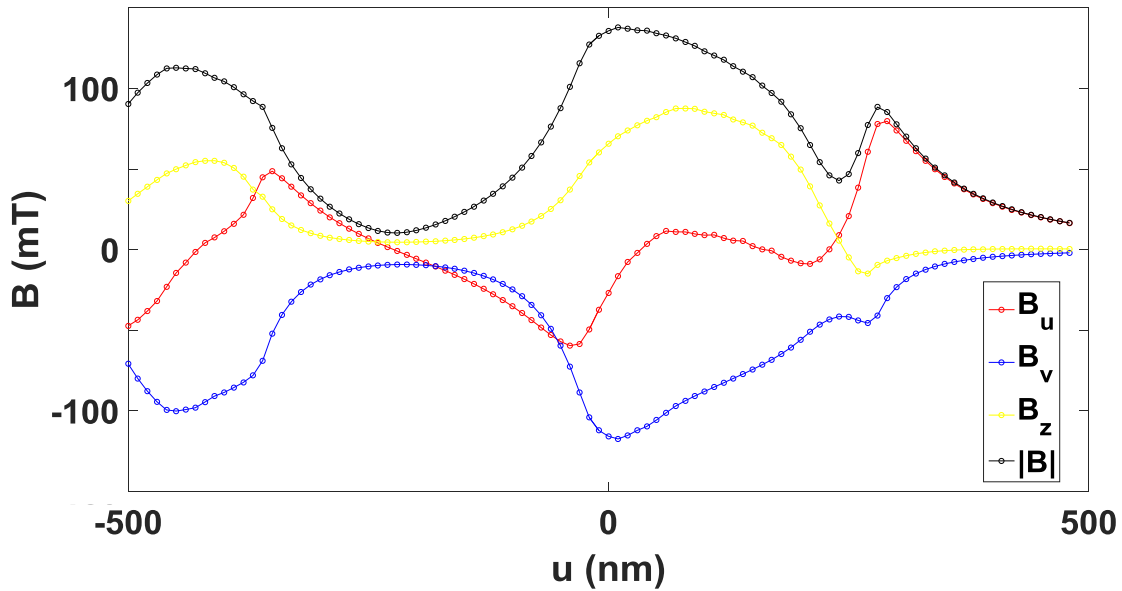

**Figure S7**. The averaged stray field calculated in the nanowire for parallel down magnetization ($B_{ext,y} = -75\ mT$) as function of the nanowire coordinate u.

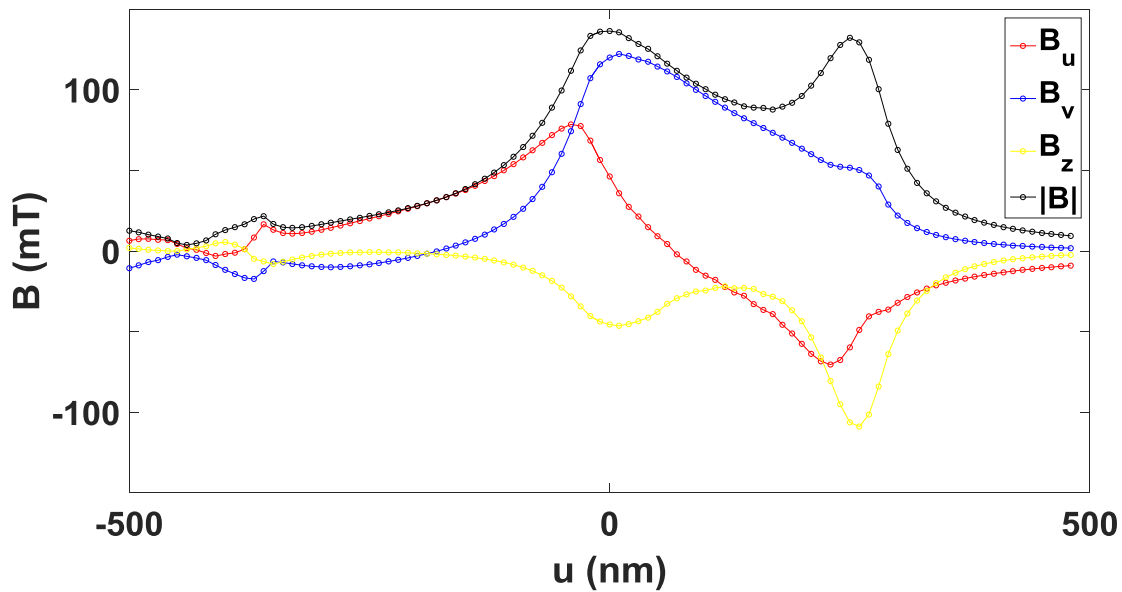

**Figure S8**. The averaged stray field in the nanowire for anti-parallel down-up magnetization ($B_{ext,y} = 50\ mT$) as function of the nanowire coordinate u.

16